\documentclass[3p, preprint]{elsarticle}

\usepackage{mathtools}
\usepackage{amsmath}
\usepackage{amssymb}
\usepackage{graphicx}
\usepackage[linesnumbered]{algorithm2e}
\usepackage{hyperref}
\usepackage{diagbox}
\usepackage{listings}
\usepackage{float}
\usepackage{placeins}
\usepackage{subcaption}
\usepackage{todonotes}
\usepackage{mathrsfs}
\usepackage{dsfont}
\usepackage{xcolor}

\usepackage{bm}

\DeclareMathOperator*{\argmin}{arg\,min}

\DeclareMathAlphabet\mathbfcal{OMS}{cmsy}{b}{n}

\newcommand{\bmeps}{\bm{\varepsilon}}
\newcommand{\epsavg}{\overline{\bmeps}}
\newcommand{\bmgam}{\bm{\Gamma}}

\newcommand{\conv}{\ast}

\newcommand{\bmx}{\bm{x}}
\newcommand{\bms}{\bm{s}}
\newcommand{\bmm}{\bm{m}}
\newcommand{\bmy}{\bm{y}}

\newcommand{\Gop}{\bm{G}}
\newcommand{\Ghat}{\widehat{\Gop}}
\newcommand{\Phiop}{\bm{\Phi}}
\newcommand{\Linop}{\mathbfcal{L}}
\newcommand{\Aop}{\Linop_a}
\newcommand{\Hop}{\mathbfcal{H}}
\newcommand{\Bop}{\mathbfcal{B}}
\newcommand{\ident}{\bm{I}}

\newcommand{\Htoep}{\bm{H}}

\newcommand{\Kroner}{Kröner}
\newcommand{\Loss}{\mathscr{L}}
\newcommand{\Ftheta}{\mathbfcal{F}_\theta}

\newenvironment{revision} {} {}

\begin{document}

\title{Recurrent Localization Networks applied to the Lippmann-Schwinger Equation}

\author{Conlain Kelly \texorpdfstring{\fnref{fn1}} {} }
\ead{ckelly84@gatech.edu}

\author{Surya R. Kalidindi  \texorpdfstring{\fnref{fn1}\corref{cor1}} {} }%
\ead{surya.kalidindi@me.gatech.edu}

\cortext[cor1]{Corresponding author}
\fntext[fn1]{Georgia Institute of Technology, Atlanta, Georgia 30332, USA}

\date{\today}

\begin{abstract}
The bulk of computational approaches for modeling physical systems in materials science derive from either analytical (i.e. physics based) or data-driven (i.e. machine-learning based) origins. In order to combine the strengths of these two approaches, we advance a novel machine learning approach for solving equations of the generalized Lippmann-Schwinger (L-S) type. In this paradigm, a given problem is converted into an equivalent L-S equation and solved as an optimization problem, where the optimization procedure is calibrated to the problem at hand. As part of a learning-based loop unrolling, we use a recurrent convolutional neural network to iteratively solve the governing equations for a field of interest. This architecture leverages the generalizability and computational efficiency of machine learning approaches, but also permits a physics-based interpretation. We demonstrate our learning approach on the two-phase elastic localization problem, where it achieves excellent accuracy on the predictions of the local (i.e., voxel-level) elastic strains. Since numerous governing equations can be converted into an equivalent L-S form, the proposed architecture has potential applications across a range of multiscale materials phenomena.

\end{abstract} 


\maketitle

\section{Introduction}\label{sec:intro}
Most problems in materials science and engineering require the exploration of  linkages between materials processing history, materials structure, and materials properties. Generally referred as Process-Structure-Property linkages \cite{kalidindiMatin}, they constitute the core materials knowledge needed to drive materials innovation supporting advances in technology \cite{mgi_plan_2014}. Traditional physics-based numerical methods have long been the standard for solving the governing field equations underpinning these linkages. For mechanical problems, these have included ubiquitous finite element methods \cite{morton_mayers_2005, roters2010_fea, mowei_2007_cahn, mcdowell2009} as well as FFT-based spectral methods \cite{moulinec1998, de_Geus_2017, michel2001_fft}. However, standard solvers can constitute a major performance bottleneck in problems which require repeated solution over varied inputs, such as inverse problems \cite{sneider2001, wu_2007_closure, jain2014} and multi-scale materials design \cite{Parno_2016, horstemeyer_2009, chen2020}. 

As an alternative, machine learning (ML) provides tools to approximate unknown linkages in a parametrized fashion, with great success in many domains \cite{carleo_2019physicsML}. One of the most successful classes of ML models is neural networks \cite{Schmidhuber_2015}, which have been applied with excellent results both in general applications \cite{carleo_2019physicsML, krizhevvsky2012_alexnet, adler2018, He_resnet}, and within materials science \cite{chen2020, yang2018Homogenization, yang2019,  Schmidt2019RecentAA}. Unfortunately, ML models tend to act as ``black boxes'' whose inner workings do not permit the depth of analysis provided by purely physics-based models \cite{buhrmester2019analysis}. There is a clear demand for approaches that leverage the advantages of both methodologies in order to build reliable, scalable, and interpretable reduced-order models. 

One example of such an effort is the Materials Knowledge Systems (MKS) framework \cite{brough2016_mks, Latypov2018MaterialsKS}. Aimed at multiscale materials design \cite{kalidindi_bayes_2019, kalidindi2019_ela}, MKS formulates the governing field equations for heterogeneous materials in a Lippmann-Schwinger (L-S) form \cite{moulinec1998, lippmannschwinger}. Using regression techniques to calibrate the first-order terms of a series expansion to the L-S equation, MKS presents a generalized approach for solving a broad class of scale-bridging problems in materials design and optimization \cite{ brough2017, priddy2017}. However, improving the accuracy of these models requires higher-order L-S terms, which rapidly become more computationally expensive.

As an alternative, we propose an approach inspired by the intersections between iterative spectral methods \cite{lebensohn2020} and recent advances in inverse imaging \cite{adler2018, putzky2017}; we cast the recurrent L-S equation as an optimization problem. Rather than employing a predefined optimization strategy, such as gradient descent or conjugate gradients \cite{Stein1952GradientMI}, the optimizer is posed as a recurrent collection of convolutional neural networks. After being calibrated to available curated data (e.g., results of FEA simulations, phase field models), these networks act as proximal (or ``update'') operators which take in a candidate solution and output an improved version. This iterative methodology emphasizes the underlying physics to permit greater model robustness and deeper analysis. 

In this paper, we begin with a brief analysis of the L-S equation and its application to the linear elasticity problem\begin{revision}, followed by discussion on the general L-S equation\end{revision}. Using this analysis, we then demonstrate how the L-S equation can be naturally posed as a machine learning problem, and how a neural network can learn proximal operations which minimize a physical quantity (e.g., stress field divergence) within a solution field. By exploring the interplay between the physical and computational interpretations of the L-S equation, we provide insight into a new class of ML models for materials science. We then analyze which aspects of our approach provide the greatest gains by exploring various model configurations, reinforcing the value of an iterative (rather than feed-forward) approach. Finally, we evaluate our methodology on the problem of elastic localization and compare it to a previous machine learning model.

\section{Background} \label{sec:background}

\subsection{Linear Elasticity and L-S}\label{subsec:ElasticBackground}

Originally developed in the context of quantum mechanics \cite{lippmannschwinger}, the L-S equation -- or class of equations -- is an implicit integral form that can represent a fairly general space of physical phenomena. The L-S equation is especially useful in the context of physics of heterogeneous media with spatially-varying physical parameters: stiffness, conductivity, density, etc. \begin{revision} We defer discussion of the general Lippmann-Schwinger form until Section \ref{subsec:generalLS}.\end{revision}

\begin{figure}
    \centering
    \includegraphics[width=0.5\columnwidth]{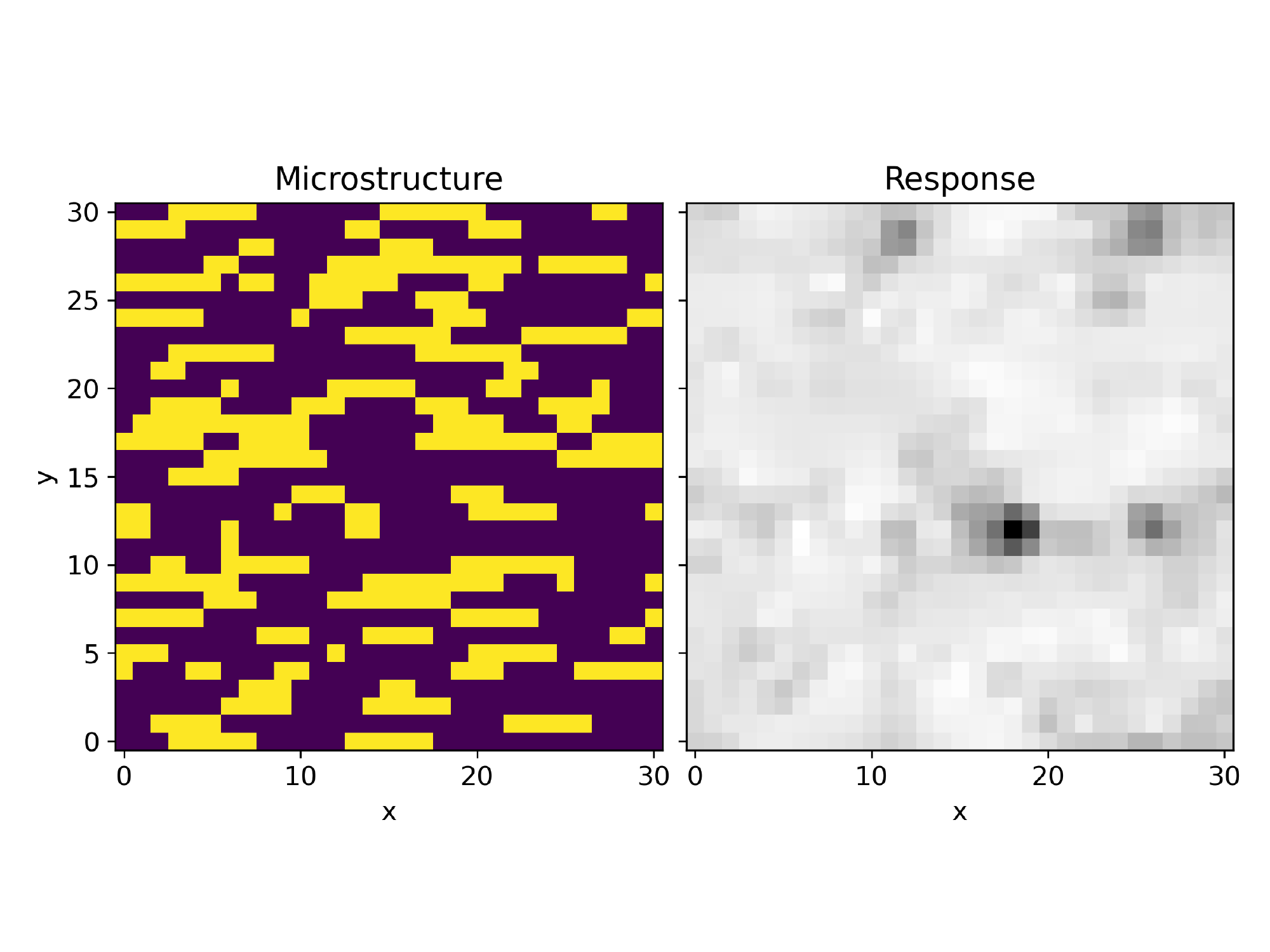}
    \caption{Example microstructure-strain field pair for two-phase elastic localization. Yellow is high-stiffness phase; purple is low-stiffness phase; contrast between elastic moduli is $CR=50$}
    \label{fig:example_io}
\end{figure}

As a case study, we consider the problem of computing the internal elastic strain field of a composite material undergoing bulk stresses or strains \cite{moulinec1998,lebensohn2020, landi2010}. The composite microstructure is assumed to be composed of two or more distinct phases (i.e., thermodynamic material constituents), each exhibiting its own constitutive laws. This problem is herein referred to as elastic localization. An example two-phase microstructure and corresponding elastic strain field are presented in Figure \ref{fig:example_io}.

Physically, elastic localization is governed by a generalized Hooke's law relating the variation in stress $\bm{\sigma}(\bmx)$, strain $\bmeps(\bmx)$, and stiffness $\bm{C}(\bmx)$ over some volume $\bm{V}$, along with the demand that the equilibrium stress field be divergence-free:

\begin{align}
    \bm{\sigma} &= \bm{C} \bmeps, \label{eq:const} \\
    \nabla \cdot \bm{\sigma} &= 0.
    \label{eq:hooke}
\end{align}
\noindent We consider constant periodic boundary conditions that correspond to the imposed volume-averaged strain, $\epsavg$. With these choices, one can model the internal mechanical response of a representative volume element (RVE) of the material. In these models, the RVE often serves as a statistical unit cell of a larger material structure. We note that the problem statement expressed here serves as a simple illustration of the L-S approach, which has been successfully applied to more complex material systems and/or boundary conditions \cite{lebensohn2020, brough2017}.

Following the work of \Kroner ~\cite{kroner1972}, this system is converted \begin{revision} into an equivalent form (the elastic L-S equation)\end{revision} by splitting the local elastic stiffness $\bm{C}$ into a selected (constant) reference value $\bm{C}^R$ and a perturbation value $\bm{C}'$. Substituting \begin{revision} these stiffness tensors into Equations (\ref{eq:const}) and (\ref{eq:hooke}) \end{revision} provides a partitioned version of the governing field equations:
\begin{align}
    \nabla \cdot (\bm{C}^R \bmeps) = - \nabla \cdot (\bm{C}' \bmeps) 
    \label{eq:Hookesplit}
\end{align}
\noindent
\begin{revision}
Observe that the left-hand side of this equation is a linear operator, denoted as $\Linop$, which acts on $\bmeps$:

\begin{equation}
   \Linop \bmeps \equiv \nabla \cdot (\bm{C}^R \bmeps) = \bm{C}^R \nabla \cdot  \bmeps
\end{equation} 

\noindent Clearly $\Linop \bmeps$ is linear in $\bmeps$; therefore, it will have a corresponding Green's function $\Gop(\bmx, \bms)$ \cite{green1828_essay}. Since divergence is uniform in space, we make the simplification $\Gop(\bmx, \bms) = \Gop(\bmx - \bms)$. This function represents the system's response to an impulse inhomogeneity  

\begin{equation}
    \Linop \Gop(\bmx - \bms) = \delta(\bmx - \bms)
\end{equation}

\noindent where $\delta$ denotes the Dirac-delta function.\end{revision} 

We also partition the strain field as $\bmeps = \bmeps^R + \bmeps'$, where $\bmeps^R$ represents a (constant) reference strain determined by boundary conditions  (this is also equal to the internal average strain $\epsavg$) and $\bmeps'$ is the corresponding perturbation strain. \begin{revision} Both $\bm{C}^R$ and $\bmeps^R$ are constant, so $\nabla \cdot (\bm{C}^R \bmeps^R) = 0$ and thus $\bmeps^R$ is the homogeneous solution to $\Linop$. For a given inhomogeniety $\bm{b}(\bmx)$, we can use the Green's function to solve for the particular solution $\bmeps'(\bmx)$:

\begin{align}
    \Linop \bmeps(\bmx) &= \bm{b}(\bmx) \\
    \implies \bmeps(\bmx) &= \bmeps^R + \int_{\bm{V}} \bm{G}(\bmx - \bms) \bm{b}(\bms) d\bms
\end{align}

Now we formally treat the right-hand side of Equation (\ref{eq:Hookesplit}) as the inhomogeneity $\bm{b}$ to obtain the elastic Lippmann-Schwinger Equation:
\end{revision}
\begin{align}
    \bmeps'(\bmx) &= - \int_{\bm{V}} \Gop(\bmx - \bms) \left( \nabla \cdot \left[ \bm{C}'(\bms) \bmeps(\bms) \right] \right) d\bms  .
    \intertext{or}
    \bmeps(\bmx) &= \epsavg - \int_{\bm{V}} \Gop(\bmx - \bms) \left( \nabla \cdot \left[ \bm{C}'(\bms) \bmeps(\bms) \right] \right) d\bms
    \label{eq:elasticLS_before}
\end{align} 
\noindent \begin{revision}
Since many governing equations can be converted into a similar form, we refer to their transformed version as an ``L-S form''; that is, Equation (\ref{eq:elasticLS_before}) is the Lippmann-Schwinger equation corresponding to the elastic governing equations. A Lippmann-Schwinger derivation corresponding to a general governing equation is presented in Section \ref{subsec:generalLS}. The reader is referred to Refs.  \cite{eisler_1969, wikipedia2020_greensfunction} for more background on Green's functions. 
\end{revision}

Using insights from previous work \cite{yang2019,kalidindi_bayes_2019,landi2010}, we modify this form to make it amenable to a learning paradigm. First, we integrate-by-parts to shift the derivative onto $\Gop$ and absorb it into a new operator, $\Ghat$. Using $\conv$ to concisely denote a convolution, we obtain 
\begin{align}
\bmeps(\bmx) &= \epsavg - \int_{\bm{V}} {\Ghat}(\bmx - \bms) \bm{C}'(\bms) \bmeps(\bms) d\bms = \epsavg - \Ghat \conv (\bm{C}' \bmeps) \label{apeq:elasticLS_after}
\end{align}
\noindent Next, we define a binary microstructure representation $\bmm^h(\bmx)$ that equals 1 if the material at point $\bmx$ is of phase (or material type) $h$, and 0 otherwise. Since each phase has its own stiffness, we can project the stiffness tensor $\bm{C}$ onto each phase: $\bm{C} = \sum_h {C}^h \bmm^h$ and likewise $\bm{C'} = \sum_h {C^h}' \bmm^h$. 
Finally, we combine the Green's function terms with the ${C^h}'$ expression to obtain yet another operator $\bmgam(\bmx - \bms)^h \equiv \Ghat(\bmx - \bms) {C^h}'$. Applying all of these modifications, the elastic L-S form becomes

\begin{equation}
\bmeps(\bmx) = \epsavg - \sum\limits_h \bmgam^h \conv \left( \bmm^h \bmeps \right) .
\label{eq:elasticLS_final}  
\end{equation}

The problem of elastic localization has thus been reduced to a single convolutional integral containing the microstructure $\bmm^h$, candidate strain field $\bmeps$, and a physics-determined stencil $\bmgam^h$. Curiously, the first two terms appear solely as an element-wise product between $\bmm^h$ and $\bmeps$. This is due to the fact that the strain field is constrained indirectly by the divergence-free condition on $\bm{\sigma}$. One also observes that all effects of $\bm{C}$ have been absorbed into $\bmgam^h$. This corresponds to the fact that $\bmgam^h$ is not unique: infinitely many choices of $\bm{C}^R$ could result in this equation. Although mathematically equivalent to the original physics (and visually more complicated), Equation (\ref{eq:elasticLS_final}) provides significant advantages for solution over large heterogeneous volumes. Several solution strategies have been explored to address elastic localization in heterogeneous material systems, resulting from different interpretations of the L-S equation.

Mathematically, $\bmgam^h$ is a set of convolutional kernels -- one for each phase -- encoding the underlying physics of the problem. Given a strain field, it computes the corresponding stresses and peels off the strain perturbation field required to minimize the stress divergence. Using this perspective, many existing models view the L-S equation as a fixed-point equation and solve it via root-finding \cite{moulinec1998}. The rate of convergence of these methods tends to depend heavily on the variation in material properties and the choice of $\bm{C}^R$ \cite{lebensohn2020}. All of these physics-based approaches require a quantitative knowledge of the Green's function $\Gop$. 

From a computer science perspective, the term $\bmm^h \bmeps$ simply represents the strain field segmented by phase (since $\bmm^h$ is a binary indicator function). Given a collection of true structure-strain pairs, one could either learn $\bmgam^h$, or some approximation, to best conserve the equality. Following this view, several ML-based elastic localization models \cite{yang2019, landi2010} have been applied to learn (non-iterative) linkages between $\bmm^h$ and $\bmeps$ by either approximating a series expansion of $\bmgam^h$, or using a neural network to map $\bmm^h$ to $\bmeps$ directly, bypassing $\bmgam^h$ completely. The disadvantage of these models is that they either truncate or ignore the underlying physics, trying to re-learn it from data. The tension between these two perspectives leaves room for a hybrid method which retains the iterative L-S structure, but uses ML to deduce the internal details of the $\bmgam^h$ operator.

\subsection{General L-S equation} \label{subsec:generalLS}
This section presents a derivation for the general L-S \begin{revision}equation\end{revision} from a generic governing equation, drawing on the work of Moulinec \cite{moulinec1998} and \Kroner ~\cite{kroner1972}. This is provided for two reasons: to provide greater intuition and background for the L-S equation, and to motivate its compatibility with ML solvers. First, we write (in conservation form) a governing differential equation controlling a field $\bmy$ which varies over space $\bmx$ spanning some volume $\bm{V}$:

\begin{equation}
\Hop(\bmy(\bmx); \bmx) = 0
\label{eq:consform}.
\end{equation}

\noindent Observe that any governing equation can be partitioned into two coupled subequations, each governing their own subsystems. First define an operator $\Aop$ which captures all linear (and spatially homogeneous) components of $\Hop$. Now define a second (possibly nonlinear) operator $\Bop$ containing the rest of $\Hop$. One obtains the earlier example of elastic localization with the substitutions $\bmy \equiv \bmeps$, $\Aop \bmeps \equiv \nabla \cdot (\bm{C}^R \bmeps)$, and $\Bop(\bmeps) \equiv \nabla \cdot (\bm{C}' \bmeps)$. Although not explicitly denoted, both $\Aop$ and $\Bop$ may contain implicit information about the solution domain's structure (terms such as $\bm{C}$ or $\bmm$).

Using these operators we can rewrite the original equation as: 

\begin{subequations}
\begin{equation}
\Hop(\bmy; \bmx) = \Aop \bmy + \Bop(\bmy; \bmx) = 0
\label{eq:cons1}
\end{equation}
or 
\begin{equation}
\Aop \bmy = -\Bop(\bmy; \bmx) \;.
\label{eq:cons2}
\end{equation}
\end{subequations}
\noindent This partitions the governing equation into two coupled systems: a linear homogeneous system permitting only ``simple'' solutions, and a nonlinear, heterogeneous system where the solution is more complicated. Before solving the complete equation, we consider the auxiliary system

\begin{equation}
\Aop \bmy(\bmx) = \bm{b}(\bmx)
\label{eq:axueq}
\end{equation}
\noindent for some inhomogeneity $\bm b$. We define the homogeneous solution to this system as $\bmy^R$, so that $\Aop \bmy^R = 0$. Note that in general, $\bmy^R$ is determined by both $\Aop$ and the relevant boundary conditions, and for some problems there may be more than one suitable $\bmy^R$. For problems with a constant solution field on the boundary, one finds that $\bmy^R = \overline{y}$, i.e.,  the reference field is the average solution everywhere.

The choice of $\bmy^R$ induces a corresponding perturbation (or ``particular solution'') $\bmy' = \bmy - \bmy^R$. Because $\bmy^R$ is annihilated by $\Aop$, note that $\Aop \bmy = \Aop \bmy'$. Since $\Aop$ is linear, it will have a Green's function $\Gop(\bmx, \bms)$, which captures the system's impulse response \cite{eisler_1969}. \begin{revision} Using this we write the particular solution to the auxiliary equation as a convolution between $\Gop$ and $\bm{b}$ and reconstruct the complete solution:\end{revision}

\begin{align}
\Aop \Gop(\bmx, \bms) &= \delta(\bmx - \bms) \\
\implies \bmy'(\bmx) &= \int_{\bm{V}} \bm{G}(\bmx, \bms) \bm{b}(\bms) d\bms \\
\intertext{or}
\bmy(\bmx) &= \bmy^R +  \int_{\bm{V}} \bm{G}(\bmx, \bms) \bm{b}(\bms) d\bms \;.
\label{eq:LSaux}
\end{align}

\noindent Now we return to Equation (\ref{eq:cons2}) and apply the auxiliary approach, this time treating the entire $\Bop$ term as our homogeneity $\bm{b}$ (and noting the attached minus sign). Plugging this into the perturbation expression for $\bmy$ gives us:

\begin{align}
\bmy &= \bmy^R - \int_{\bm{V}} \bm{G}(\bmx, \bms) \Bop(\bmy(\bms); \bms) d\bms .
\label{eq:generalLS} 
\end{align}

\noindent This is the \begin{revision}general\end{revision} Lippmann-Schwinger equation; \begin{revision} since this derivation holds for any operator $\Hop$, we use the term ``L-S form'' for a given $\Hop$ to describe the result of partitioning and substituting that $\Hop$ into Equation (\ref{eq:generalLS}). Referring back to the example of elastic localization, Equation (\ref{eq:elasticLS_before}) is the equivalent L-S form for Hooke's law (Equation (\ref{eq:hooke})).\end{revision} Note that the linear system $\Aop$ only enters the L-S equation through the definitions of $\bmy^R$ and $\Gop$. For example, if one used the trivial choice of the identity for $\Aop$, the corresponding Green's function would just be the Dirac delta function, and Equation (\ref{eq:generalLS}) would simplify to the original governing equation.

For the L-S form to be advantageous over $\Hop$, $\Aop$ must capture a non-trivial amount of the underlying equation. There are three primary factors which make the L-S equation useful. First, it is \textit{partitioned}: the original system is broken into two coupled physical systems. This makes it similar to a traditional ``splitting method'' where the linear and nonlinear components are separated, allowing the solution of the nonlinear components to be informed by the homogeneous, linear solution. The coupling between systems means that the L-S equation is also \textit{recursive}: the presence of $\bmy$ in the inhomogeneity term leads to its appearance on both sides of the equation. If one desires to solve the problem analytically, the L-S form is likely no more useful than the original governing equation. However, the implicit structure is very suitable for iterative and optimization-based solvers \cite{michel2001_fft,lebensohn2013_crystal}. Finally, the L-S equation is \textit{convolutional}: the integral is actually a convolution between the $\Bop$ term and a (possibly-unknown) Green's function $\Gop$.
Roughly speaking, Equation (\ref{eq:generalLS}) presents the solution field $\bmy(\bmx)$ as a balance between the global homogeneous ``pull'' ($\bmy^R$) and the localized ``tug'' ($\bmy'(\bmx)$) of the solution values in a neighborhood near $\bmx$. In situations where $\Bop$ is a purely differential operator (such as  elastic localization), and with appropriate boundary conditions, Equation (\ref{eq:generalLS}) can be integrated-by-parts to shift part of $\Bop$ onto $\Gop$. This can simplify the integral term so that all of the physics is contained in a single convolutional stencil.

\subsection{Neural Networks Background}\label{subsec:MLBackground}

As one of the most popular ML tools in use, neural networks are a class of parametrized function approximators that can be calibrated to curated data \cite{Schmidhuber_2015}. At a high level, an artificial neural network (ANN) operates as an alternating sequence of tunable linear transforms and nonlinear activation functions -- mimicking the operation of physical neurons in the brain \cite{mccullock1988_neuralnets, rosenblatt1957perceptron}. Under certain conditions, a sufficiently large neural network can be shown to act as a universal function approximator \cite{Cybenko1989ApproximationBS, Pinkus1999ApproximationTO}, motivating their use in a myriad of disciplines.

Two relevant specializations are the convolutional neural network (CNN), which uses convolution with a fixed-width stencil as its transform operation \cite{lecun1998_CNNs,krizhevvsky2012_alexnet}, and the recurrent neural network (RNN), which operates on sequential data and considers latent information carried across input iterations \cite{lipton2015rnn}. These tools can be combined to model the underlying structure of various problems. A well-designed ML model trained on sufficient data can be significantly faster than an analytical equivalent and still provide reasonable accuracy \cite{streib_2020}. This comes at the expense of interpretability -- they return a ``black-box'' model which is difficult to understand and analyze \cite{buhrmester2019analysis}. Additionally, the topology of these networks (e.g., number of layers, nodes per layer, activation functions) strongly determines their success \cite{Ojha_architecture}, and the ``best'' configuration is problem-dependent and often constructed ad-hoc.

Recently there has been tremendous interest in the application of neural networks to mathematical problems \cite{carleo_2019physicsML}. Specifically, variations of recurrent CNNs have been explored \cite{andrychowicz2016learning} to learn Bayesian priors for image denoising \cite{putzky2017} or proximal operators for medical imaging \cite{adler2018}. These image analysis methods pose the problem such that the desired output is obtained via a learned optimization procedure, where the optimizer itself is formulated as a neural network. Surprisingly, these methods often employ very simple network designs, especially compared to deeper and more elaborate structures found in mainstream ML \cite{He_resnet, Iandola_squeezenet}. 

\section{Methodology}\label{sec:methodology}
\subsection{L-S as learned optimization}\label{subsec:general_math}

We now explore how the perturbation expansion and L-S form allow a governing equation to be interpreted naturally as a machine learning problem. We first define a new operator $\Phiop$ representing the entire right-hand side of Equation (\ref{eq:generalLS}). We also use $\bmm$ to represent a problem domain's underlying microstructure, which influences the inhomogeneity $\Bop$. Given a sample microstructure $\bmm^*$, we obtain the corresponding strain $\bmy^*$ by minimizing the error (or loss) $\Loss$ between $\bmy$ and $\Phiop(\bmy, \bmm^*)$ over all $\bmy$.

\begin{align}
    \Phiop(\bmy, \bmm) &\equiv \bmy^R - \int_{\bm{V}} \bm{G}(\bmx, \bms) \Bop(\bmy(\bms); \bms) d\bms \\
    \bmy^* & = \Phiop( \bmy^*; \bm{m}^* ) = \argmin_{\bmy} \Loss\left( \bmy, \Phiop( \bmy; \bm{m}^* ) \right) \label{eq:var}
\end{align}

Although $\Phiop$ may not be linear itself, linear analysis methods provide a useful interpretation: for a given microstructure $\bmm^*$, $\Phiop$ has a (possibly non-unique) generalized eigenfunction $\bmy^*$ with unit eigenvalue. Issues regarding the solution's uniqueness and existence can arise from the original governing equation's nonlinearity. In the case of the elasticity problem, the governing equation is linear, so a unique solution will exist.

Now the original problem of solving the governing equation $\Hop$ has been reduced to that of minimizing the \begin{revision}
scalar
\end{revision} loss $\Loss$ given a particular microstructure via a learned optimization strategy. To do this, we define a parametrized learner $\Ftheta$ that performs a sequence of proximal operations which progressively improve the solution field to match the governing physics. Here $\theta$ represents all possible parameters of this learner. The optimal set of parameters $\theta_{opt}$ is obtained by minimizing the expected error produced by $\Ftheta$ w.r.t. $\theta$; in the case of a CNN this represents the optimal network weights, and can be obtained via standard backpropagation \cite{rumelhart1988_backprop}. Given a microstructure and initial guess $\bmy_0$, we want $\Ftheta$ to provide a solution field which is approximately the true solution field $\bmy^*$:

\begin{align}
    \Ftheta(\bmy_0, \bmm^*) = \hat{\bmy} \approx \bmy^*
\end{align}

\begin{figure}
\centering
    \begin{subfigure}[b]{0.6\columnwidth}
        \begin{lstlisting}[mathescape=true, frame=single]
        def $\Ftheta(\bmy^R, \bmm)$:
            $\bmy_0 = \bmy^R$
            for $i = 1$ to $N$:
                $\bmy_{i} = \bmy^R + f_{i} (\bmy_{i-1}, \bmm)$
            return $\bmy_N$
        \end{lstlisting}
    
    \caption{Pseudocode for \begin{revision} optimization procedure\end{revision}}
    \label{lst:pseudocode}
    \end{subfigure}
    
    \begin{subfigure}[b]{0.6\columnwidth}
        \centering
        \includegraphics[width=\columnwidth]{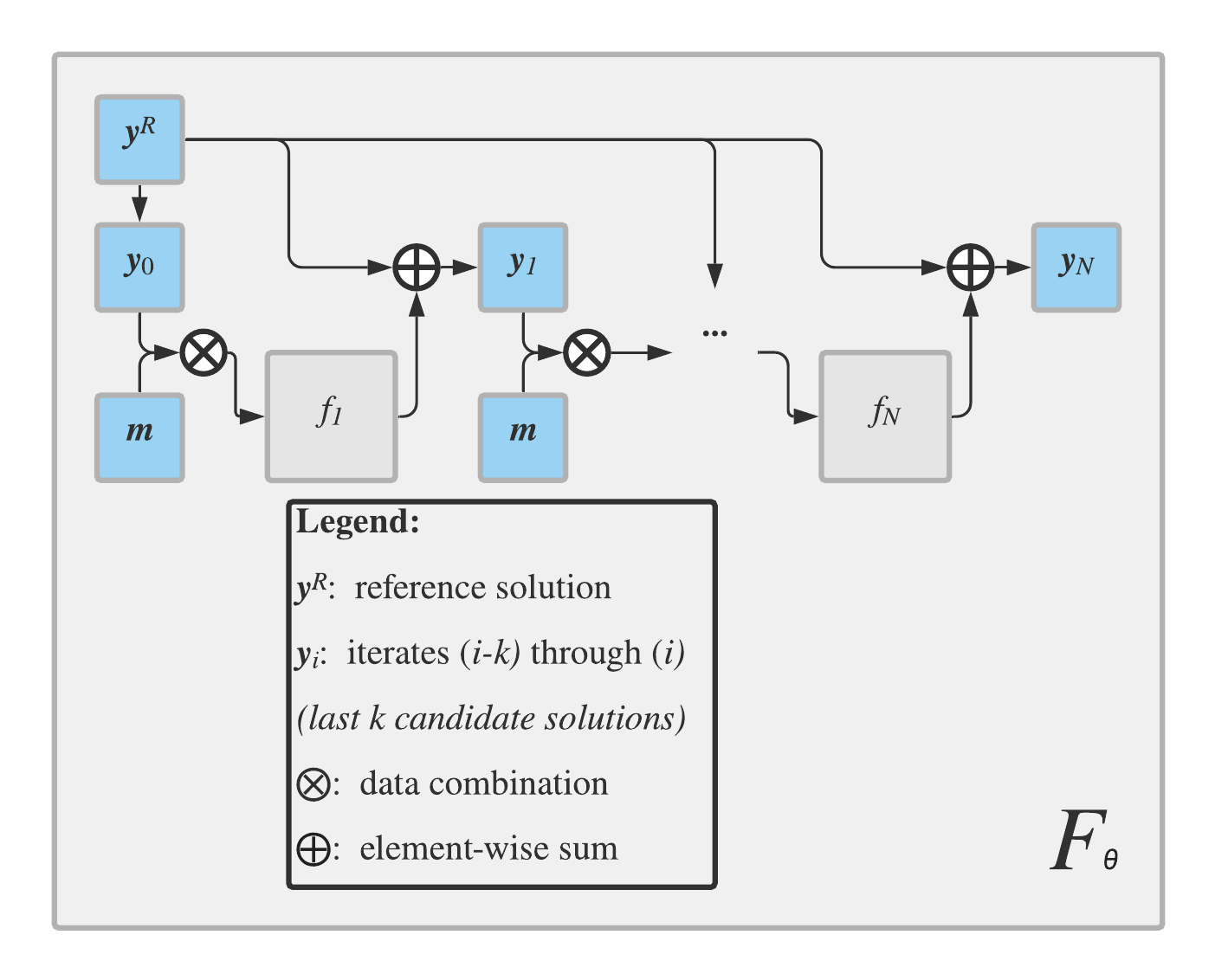}
        \caption{Visualization of data flow through \begin{revision} optimization procedure\end{revision}}
        \label{fig:RLN_diag}
    \end{subfigure}

\caption{Pseudocode and visualization for \begin{revision} optimization-based solution of the L-S equation\end{revision}}
\label{fig:RLN_overview}

\end{figure}

\noindent This is accomplished by a sequence of updates

\begin{equation}
    \bmy_{i} = \bmy^R + f_{i} (\bmy_{i-1}, \bmm^h) 
\end{equation}
\noindent where $f_i$ represents the perturbation proximal operator used at iteration $i$; given a microstructure and candidate strain field, it outputs the perturbation component corresponding to an improved estimate of the true strain field. A pseudocode and visual representation of the approach developed in this work are presented in Figure \ref{fig:RLN_overview}. Ideally, after the model is trained, $\Ftheta$ and $\Phiop$ have the same eigenfunctions (i.e., $\Ftheta(\bmy^*, \bmm^*) = \bmy^*$). 

The optimization strategy employed by the $\Ftheta$ model is directly determined by the choice of $f_i$. To explore what this might look like for the elastic problem,  consider an L2 loss function and plug in the elastic LS formula (Equation (\ref{eq:elasticLS_final})):

\begin{align}
    \Phiop( \bmy; \bmm^h ) &= \bmy^R - \sum\limits_h \bmgam^h \conv \left( \bmm^h \bmy \right) \label{eq:elastic_LS_preloss}\\
    \Loss\left( \bmy, \Phiop( \bmy; \bmm^h ) \right) &\equiv \frac{1}{2} \left( \bmy - \Phiop( \bmy; \bmm^h ) \right)^2 = \frac{1}{2} \left (\bmy - \bmy^R + \sum\limits_h \bmgam^h \conv ( \bmm^h \bmy ) \right) ^2
    \label{eq:elastic_loss}
\end{align}
\noindent The original fixed-point approach of Moulinec \cite{moulinec1998} corresponds to the choice 
\begin{equation}
    f_i^{M}( \bmy_{i-1}, \bmm^h) = - \sum\limits_h \bmgam^h \conv \left( \bmm^h \bmy_{i-1} \right) \quad \forall i 
\end{equation}

\noindent As an alternative, we can obtain a ``steepest descent'' formula by taking the gradient \cite{Stein1952GradientMI} of Equation (\ref{eq:elastic_loss}):

\begin{align}
f_i^{G}(\bmy_{i-1}, \bmm^h) &= \bmy_{i-1} - \bmy^R - \gamma_i \frac{\partial} {\partial \bmy_{i-1}} \Loss \left( \bmy_{i-1}, \Phiop( \bmy_{i-1}; \bmm^h)  \right) \\
\intertext{or}
f_i^{G}(\bmy_{i-1}, \bmm^h) &= \bmy_{i-1} - \bmy^R - \gamma_i \left(\ident + \frac{\partial} {\partial \bmy_{i-1}}  \sum\limits_h \bmgam^h \conv ( \bmm^h \bmy_{i-1} ) \right) \left( \bmy_{i-1} - \bmy^R + \sum\limits_h \bmgam^h \conv ( \bmm^h \bmy_{i-1} ) \right)
 \label{eq:gradient}
\end{align}
\noindent where $\gamma_i$ denotes the step size at iteration $i$ and $\ident$ represents the identity operator. By flattening everything into vectors and representing the convolution with an equivalent Toeplitz matrix \cite{gray_toeplitz}, one can convert the product term in Equation (\ref{eq:gradient}) into a single linear operator $\Htoep$ acting on $\bmy_{i-1}$. The linearity of $\Htoep$ comes from the fact that the $(\ident + \frac{\partial} {\partial \bmy_{i-1}} \ldots)$ term is actually independent of $\bmy_{i-1}$. Using a constant $\lambda_i$ to collect remaining terms of $\bmy^R$, the steepest descent rule becomes:

\begin{equation}
    f_i^{G}(\bmy_{i-1}, \bmm^h) = (1 - \gamma_i)(\bmy_{i-1} - \bmy^R) + \gamma_i\Htoep \bmy_{i-1} + \lambda_i \bmy^R
\end{equation}
\noindent Effectively, the gradient descent rule says that the new perturbation field is obtained by correcting the previous perturbation field ($\bmy_{i-1} - \bmy^R$) using a set of convolutional operations involving $\bmgam^h$ and $\bmm^h$, then subtracting off a factor of $\bmy^R$ such that the output perturbation field is zero-mean.

A variety of more complicated update rules have been proposed to accelerate the solution of different forms of the L-S equation \cite{lebensohn2020}. From the examples above one sees that any update rule will involve various terms of $\bmgam^h$, which itself contains both physical properties (${C^h}'$) as well as derivatives of the Green's function ($\Ghat$); more complicated update rules will simply require higher-order combinations. Therefore, if we hope to learn $f_i$, it must be parametrized in a way such that it can capture global convolutional stencils, as well as various derivatives thereof. Finally, we note that although most analytical approaches employ the same operator for each iteration, the $\Ftheta$ formulation permits varying $f_i$ across iterations. 

\subsection{CNNs for Lippmann-Schwinger}\label{subsec:CNN_method}
Most analytical minimization procedures are (by design) problem-agnostic; this means that they can be expected to work reasonably well for many problems, but may not be optimal for the problem at hand. Rather than using a predefined optimization strategy, we formulate each $f_i$ as a CNN that learns a proximal operator mapping a given microstructure and candidate solution field to an improved perturbation field. The central motivation behind using a CNN proximal operator is that given sufficient parameterization, it can emulate almost any optimization strategy; furthermore, that strategy will be customized to the problem at hand during training. Of course, this comes with the immense caveat that, absent any advances in theoretical ML, a learned optimizer will not have any provable convergence guarantees. The means that even as $N\rightarrow \infty$, our learned model may not converge to the true solution field for a given microstructure. In practice, however, the model can be tuned and trained until it consistently produces solutions within acceptable error tolerances. 

We define the $\Ftheta$ model imbued with a CNN $f_i$ as a recurrent localization network (RLN), which performs the following operations during both training and evaluation phases: given a microstructure $\bmm$ and initial guess $\bmy^R$, estimate the true solution field by refining it over $N$ iterations. At iteration $i$, the microstructure is combined with candidate solutions from previous iterations and passed through CNN $f_i$. Note that $f_i$ outputs a \textit{perturbation} field $\bmy_i'$. The reference solution $\bmy^R$ is already known, so there is no need to learn that. In order to simulate a multi-step solver \cite{morton_mayers_2005, ullah_multistep} \begin{revision}
and estimate higher-order derivative terms\end{revision}, $f_i$ considers the last $k$ solutions via multiple input channels (rather than just $y_{i-1}$). \begin{revision}
One could potentially achieve this property, and perhaps obtain better results, by using GRU or LSTM modules \cite{putzky2017} which learn a ``latent'' state to pass between iterations; however, 3D convolutional implementations for these operations were not part of major ML libraries at the time of writing.
\end{revision}

Specifically for elastic localization, $\bmm$ and $\bmy$ are combined via element-wise multiplication following Equation (\ref{eq:elasticLS_final}). To enforce computational stability, all strains are normalized by the average strain $\epsavg$. Moreover, the output of each $f_i$ network has its average subtracted to enforce the constraint that the perturbation strain field is always zero-mean. 

Following prior work \cite{adler2018}, we define a full RLN as using a different $f_i$ for each iteration (although we use the same CNN structure for each), for a total of $N$ distinct networks. The idea behind this is to allow the network capture different properties at each iteration, akin to terms in a series expansion. Having significantly more tunable parameters, this configuration provides the most expressive model. By allowing different operators to be employed at different iterations, this approach also deviates the most from standard analytical optimization procedures. 

Alternatively, one may wish to reuse the same intermediate network across iterations ($f_{i} = f ~ \forall i$). This approach is denoted as RLN-t since the weights are \textit{tied} between iterations.  This means that the same operator will be employed at each iteration; however, since a time series is fed into each $f_i$ (rather than a single data point), the RLN-t is still able to learn higher-order derivatives. It has the primary advantage of simplicity and efficiency, since it uses a factor of $N$ fewer parameters than the full RLN. 

Finally, we test the importance of the iterative and recurrent nature by considering a single $f_i$ network, i.e., choosing $N=1$. We call this a feed-forward localization network (FLN) and use it as a control to quantify the benefits of iteration \textit{vs.} network design. Although the RLN-t and the FLN have the same number of parameters, the RLN-t uses each parameter $N$ times, effectively simulating a deeper network. 

For $N > 1$, the proximal CNNs are all calibrated simultaneously via backpropagation. During training a batch of true structure-strain pairs are fed through $\Ftheta$ in its entirety, and all proximal networks are updated simultaneously to minimize a \begin{revision}
calibration\end{revision} loss function $\Loss^{(cal)}$ (different from the solution field loss $\Loss$ above). Rather than only consider the loss of the last iterate, we use a weighted sum of the loss of individual iterates: $\Loss^{(cal)} = \sum_i w_i \Loss^{(cal)}_i$ for some weights $w_i$. The goal is to encourage the network to progress between iterations, while also finding the best possible solution. \begin{revision} Following the analysis of Andrychowicz et al. \cite{andrychowicz2016learning} we interpret the use of $\Loss^{(cal)}$ as a variant of Backpropogation Through Time \cite{Mozer1989_backprop_time}. The choice of weights $w_i$ could theoretically act as a form of stabilization or even regularization. By requiring that each iteration output a reasonable candidate solution, each proximal operator is constrained to behave somewhat physically, which might help prevent overfitting. However, this means that the network is encouraged to make larger changes in early iterations, potentially reducing its final-iterate accuracy. Conversely, if only the final result is important, then intermediate iterations could explore the loss curve more. However, only updating based on the last iteration will slow the model's training, and possibly increase the chances of the network weights converging to a poor local minimum. Clearly further experiments are required to explore these hypotheses.\end{revision}

Following similar works \cite{putzky2017} we chose the uniform weighting $w_i = 1 ~ \forall i$. This induces the network to make larger corrections in early iterations (to avoid carrying costly errors through several iterations) and relatively smaller corrections in later iterations. However, our numerical experiments (Section \ref{sec:experiments}) indicate that perhaps a different weighting might help the network capture fine-scale microstructure features. The appropriate number of iterations depends on the specific problem; for elasticity $N=5$ proved sufficient in both the RLN and the RLN-t, and more iterations yielded little benefit. \begin {revision} For the number of previous iterations to track, the value $k=2$ was chosen. The above choices of hyperparameter were largely heuristic and made for simplicity, but they worked well for elasticity; for a more intensive problem a cross-validation procedure would be a better method for their selection.\end{revision}

\subsection{Proximal operator design}
\begin{figure}
    \centering
    \includegraphics[width=0.5\columnwidth]{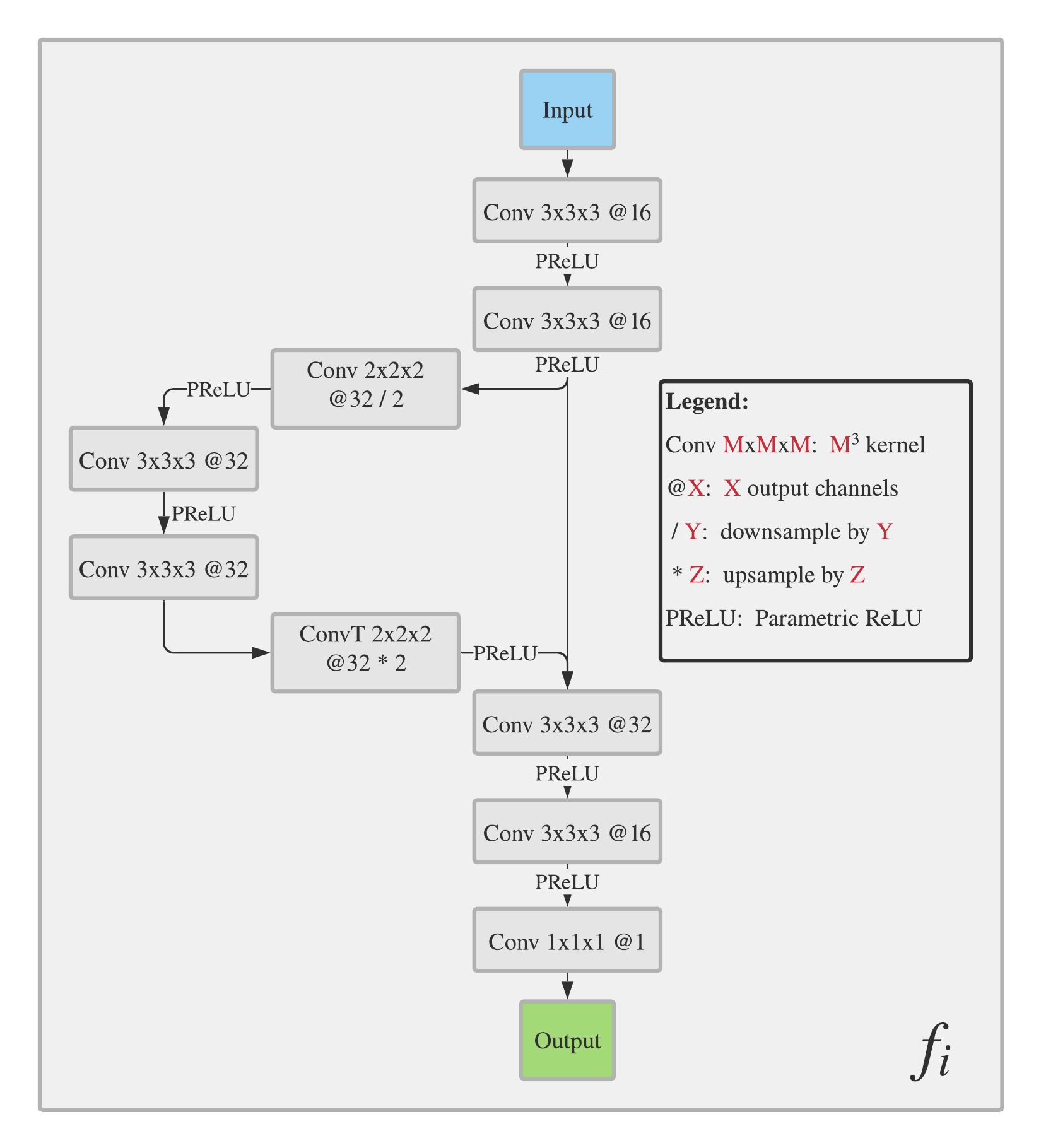}
    \caption{Architecture for internal $f_i$ networks}
    \label{fig:fi_arch}
\end{figure}

Various network \begin{revision} topologies\end{revision} for $f_i$ \cite{adler2018, putzky2017, milletari2016vnet} were tested with one goal in mind: use convolutional kernels to effectively capture local interactions. The architecture that proved most successful for elasticity was based roughly on a V-Net \cite{milletari2016vnet} and is presented in Figure \ref{fig:fi_arch}. By combining information across length scales through the use of up- and down-sampling operations, this architecture is able to encode and combine both fine- and coarse-grained features. Notably, even a simple sequence of 3 convolutional layers (similar to that of Adler \cite{adler2018}) worked reasonably well. \begin{revision}As with the hyperparameters above, a cross-validation procedure could aid in picking a superior network topology for a given problem; for this work, we intentionally avoided hyperparameter optimization to emphasize the mathematical analysis and iterative approach.\end{revision}

For this study, most convolutional kernels are 3x3x3, with 1x1x1 kernels used for channel reduction and 2x2x2 for up/down-sampling, summing to $\approx$ 160,000 learnable parameters for each proximal network $f_i$. A parametric rectified linear unit (PReLU) activation function \cite{he2015_prelu} was used was used after the 3x3x3 convolutions. This activation is similar the regular ReLU, but has a non-zero slope $\alpha$ for negative inputs; $\alpha$ is trained with all the other network weights during backpropogation. In our experiments this improved the model's stability during training and accuracy during testing. Effectively, it allows the network to be much more expressive with only a few extra parameters -- changing one $\alpha$ scales the entire output of a channel, rather than $3 \times 3 \times 3=27$ weights representing each voxel in that channel's filter. 

\section{Linear Elasticity Experiments}\label{sec:experiments}

We now present results from the application of RLN-type models to the elastic localization problem in two-phase composites. A synthetic dataset of 20,480 $31 \times 31 \times 31$ microstructure/strain field pairs was generated and randomly split 40\% / 20 \% / 40\% into \begin{revision}disjoint\end{revision} train/validation/test sets, respectively. The train set was used to calibrate $\Ftheta$, the validation set served to measure its quality while training, and the test set was used to compute error metrics.

The microstructures were generated via PyMKS \cite{brough2016_mks} by creating a uniform random field, applying \begin{revision}a set of Gaussian filters\end{revision}, and thresholding the result. In this microstructure generation process, the filter's width in each direction controls the characteristic size and shape of the material grains, and the threshold controls the relative volume fraction of each phase. 

\begin{figure}
    \centering
    \includegraphics[width = \columnwidth]{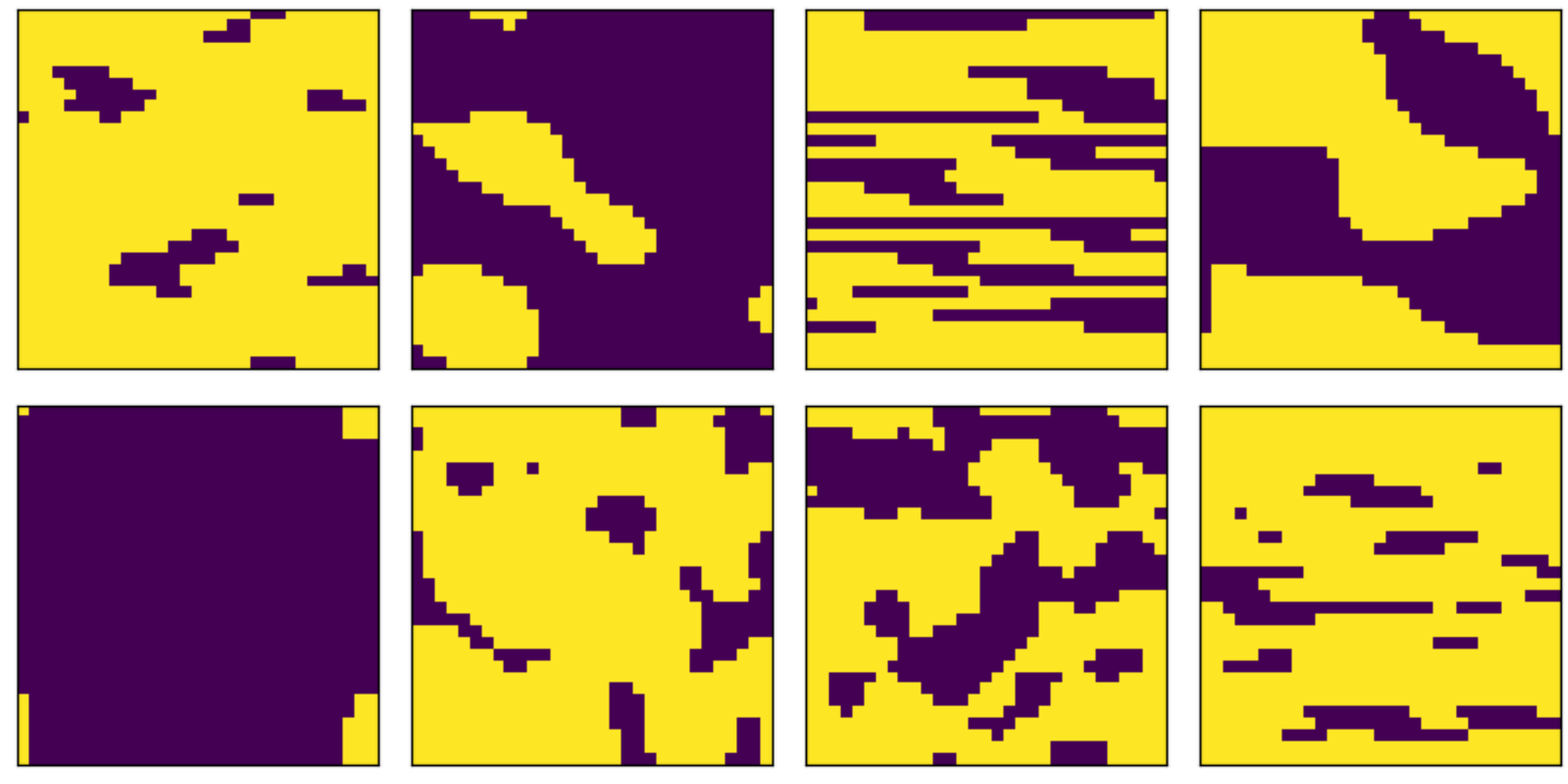}
    \caption{Random sample of 8 training microstructures sliced along $(y,z)$ axis. Yellow is high-stiffness phase, purple is low-stiffness.}
    \label{fig:micro_sample}
\end{figure}

Combinations of these four parameters were generated via a Latin hypercube sampling procedure to generate 4096 design points for microstructure generation. For each design point, 5 random microstructures were generated in order to produce statistically similar inputs, so that the datasets had some statistical redundancy. A random sample of 8 training microstructures is displayed in Figure \ref{fig:micro_sample}.  
\begin{revision}
By varying these design parameters, one can cover a vast subspace of all possible microstructures, although we note that the space of all microstructures in intractably large -- ignoring circular symmetries there are $O(2^S)$ possible different microstructures with $S$ voxels. By selecting our training and testing sets using the same procedure, we demonstrate the RLN's ability to interpolate between ``familiar'' microstructure samples, but not to extrapolate to microstructures which lie outside the closure of the training set.\end{revision}
It is important to note that these microstructures do not necessarily correspond to thermodynamically favorable structures (i.e. they might appear in nature). However, this is a limitation only in data and not methodology -- by using a fully-convolutional structure \cite{long2015fully} the RLN can handle any voxelized input, including experimental data. 

The elastic strain fields in each microstructure were obtained from previously-built finite element models \cite{yang2019}. One major simplification is that even though the imposed 3D strain field is actually a second-order tensor (and the elastic stiffness field a fourth-order tensor), the linear nature of the problem allows us to solve for each component individually and superimpose their solutions as needed \cite{landi2010}. As such we apply an overall displacement to a material RVE along only the $x$-axis and we focus solely on the $\epsilon_{xx}$ term. For these simulations, a total strain of 0.1\% was applied via periodic boundary conditions. 

The relevant material parameters to be prescribed are thus the elastic moduli $E_1, E_2$ and Poisson ratios $\nu_1, \nu_2$ of the two phases. To roughly match most common metals, we chose $\nu_1 = \nu_2 = 0.3$. With these selections, the contrast ratio $CR \equiv \frac{E_2}{E_1}$ has the most dominant role on the final strain field. With the choice $E_2 \geq E_1$, one observes that $CR \geq 1$. In general, as the contrast in stiffnesses increases, the problem becomes harder to solve with both iterative and data-driven methods \cite{lebensohn2020}. Following prior work \cite{yang2019}, we tested the RLN on contrast ratios of 10 and 50. 

\begin{figure}
    \centering
    \includegraphics[width=\columnwidth]{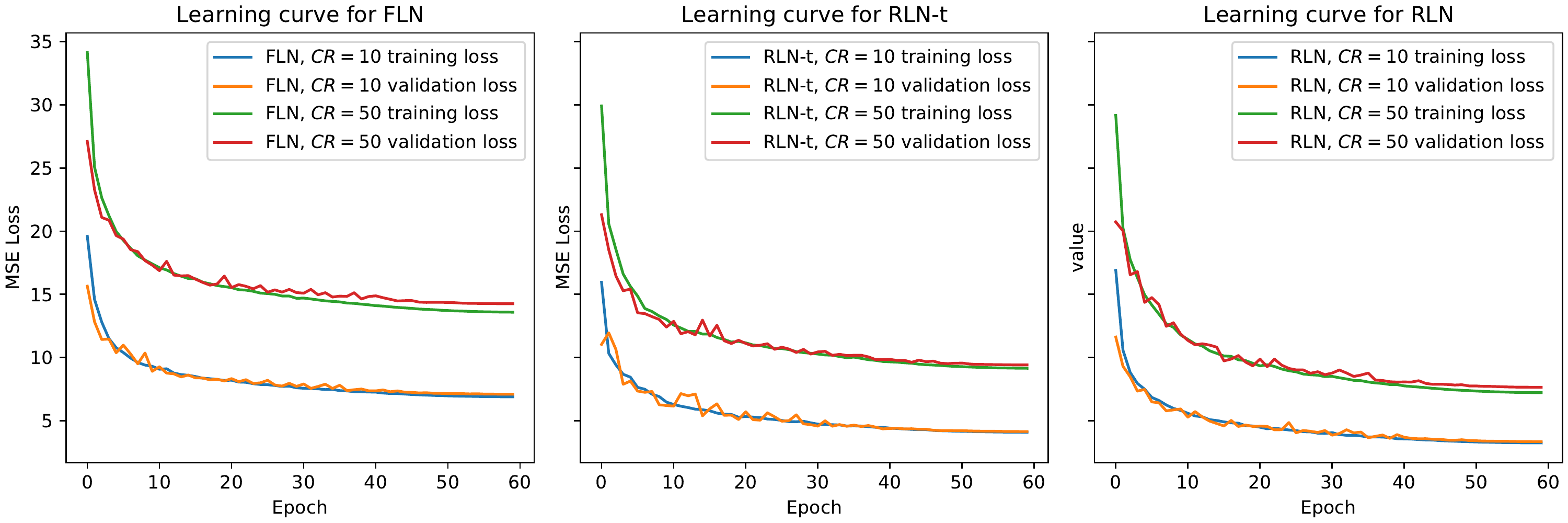}
    \caption{\begin{revision}Learning curve containing calibration losses for each model and contrast ratio.\end{revision}}
    \label{fig:learning_curve}
\end{figure}

Each RLN model was implemented in PyTorch \cite{Paszke_2019pytorch} and trained independently for 60 epochs on an NVIDIA V100 GPU \cite{PACE}, which took approximately 8 hours. The RLN models were all calibrated using the Adam optimizer, a Mean Square Error loss, and a Cosine annealing learning rate decay \cite{loshchilov2017sgdr}. The last choice proved especially important since the models demonstrated great sensitivity to learning rate; introducing the cosine decay helped the model converge smoothly and quickly. After training, the epoch with the lowest validation loss was chosen as the ``optimal''. 
In reality, the training and validation losses were tightly coupled \begin{revision} as demonstrated in Figure \ref{fig:learning_curve}. Note that training losses are aggregated during the epoch, whereas validation losses are computed after the epoch is complete, which causes the validation loss to occasionally appear lower.\end{revision} 

\subsection{Results}

\begin{figure}
    \centering
    \includegraphics[width=0.5\columnwidth]{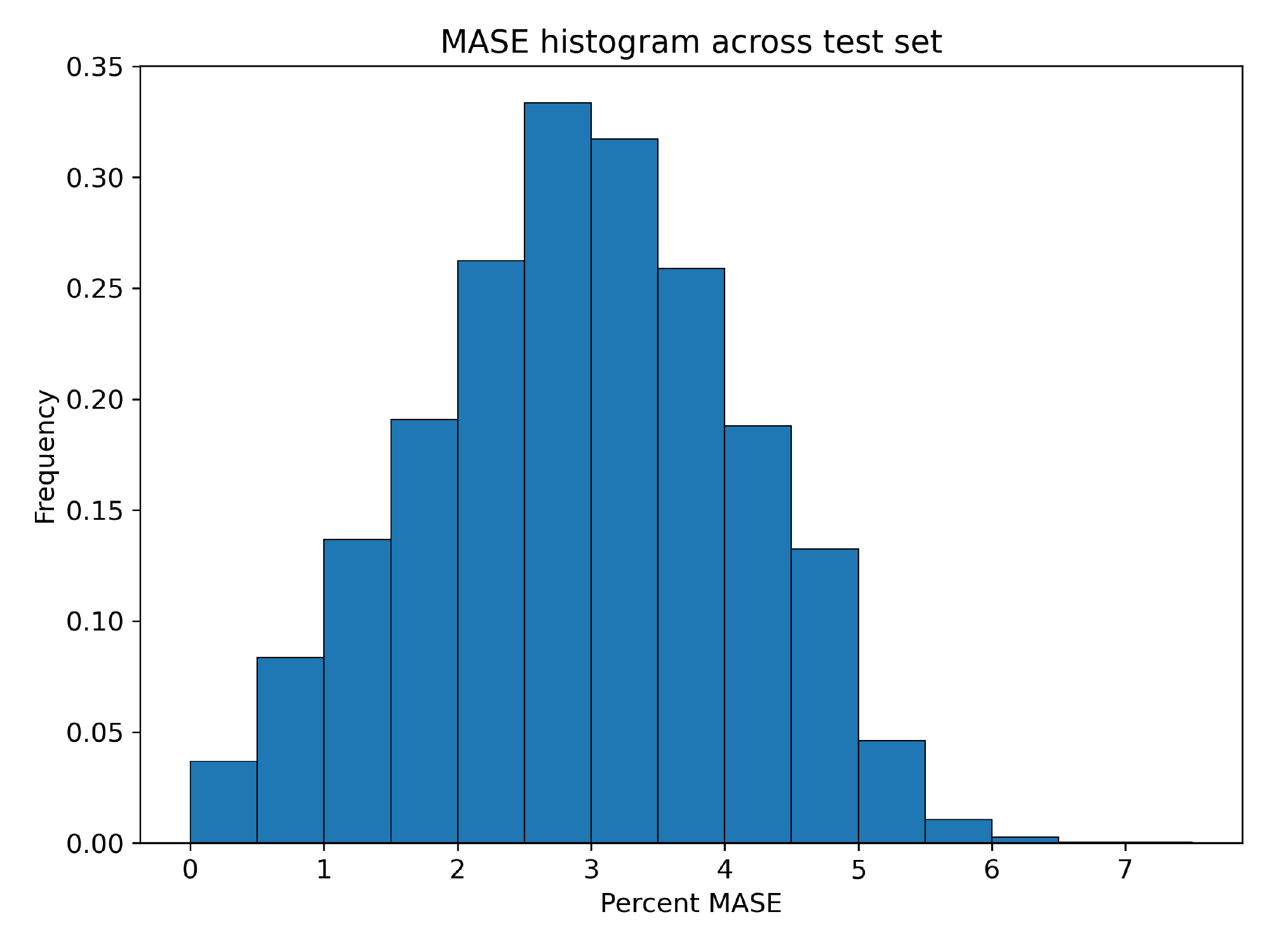}
    \caption{MASE distribution for RLN, $CR=50$}
    \label{fig:errhist}
\end{figure}

Following Yang et al. \cite{yang2019}, the accuracy of the model was evaluated for each instance using the mean absolute strain error (MASE) \begin{revision} over all $S$ voxels\end{revision}:
\begin{equation}
    MASE(\bmy_{pred}, \bmy_{true}) = 100 \times \frac{1}{S} \sum\limits_{i=1}^S\left| \frac{\bmy_{pred}[i] - \bmy_{true}[i]} {\bmy^R} \right|
    \label{eq:ASE}.
\end{equation}
\noindent Figure \ref{fig:errhist} presents the MASE distribution across each microstructure/strain pair in the test set. Note that the MASE is an aggregate measure which measures RVE-wide error; the pointwise error variation within a microstructure is explored below. The mean and standard deviation of the MASE distribution are collected in Table \ref{tab:res10} for the RLN-type models. For comparison we also present results from a recent study \cite{yang2019} using a feed-forward deep learning (DL) model to predict the strain fields; note that the DL model was trained and tested on its own dataset prior to this effort. 

\begin{table}
\centering
\begin{tabular}{|p{0.45\columnwidth}|p{0.45\columnwidth}|}\hline
\multicolumn{1}{|c|}{\normalsize \bf Model} & \multicolumn{1}{c|}{\normalsize \textbf{MASE} (mean $\pm$ std. dev.)}                          \\ \hline \hline
\multicolumn{2}{|c|}{ \bf Contrast-10}                               \\ \hline
Comparison DL model \cite{yang2019}      & 3.07\%$\pm$1.22\%     \\ \hline
FLN                                     & 4.98\%$\pm$1.49\%     \\ \hline
RLN-t                                   & 1.81\%$\pm$0.58\%     \\ \hline
RLN                                     & 1.21\%$\pm$0.37\%     \\ \hline \hline
\multicolumn{2}{|c|}{ \bf Contrast-50}                               \\ \hline
Comparison DL model                     & 5.71\%$\pm$2.46\%       \\ \hline
FLN                                     & 9.23\%$\pm$3.29\%       \\ \hline
RLN-t                                   & 4.26\%$\pm$1.65\%       \\ \hline
RLN                                     & 2.92\%$\pm$1.17\%       \\ \hline
\end{tabular}
\caption{Test-set MASE metrics for RLN and control models, for both $CR=10$ and $CR=50$}
\label{tab:res10}
\end{table}

\begin{revision}
It is important to note that the DL model had a fundamentally different architecture: it was designed to work on $21^3$ voxel structures (whereas ours was tested on $31^3$), and it predicted the strain one voxel at a time (whereas ours predicts strain across the entire microstructure simultaneously). The dataset for the DL model employed large amounts of data augmentation using circular permutations, and contained significantly more input-output pairs. Finally, the DL model employed linear layers (to collapse to a single output); the size of these layers implies that the DL model used substantially more network weights than the RLN. As a result, it is difficult to compare the DL dataset and results with ours. We emphasize that the DL model represents a strong comparison model for ML elastic localization, but that objective ranking of relative performance is difficult. Nevertheless, the RLN architecture is able to produce significantly more accurate strain field estimates on the RLN dataset than the DL architecture produces on the DL dataset. Keeping these caveats in mind, the following analysis explores the difference between RLN configurations.
\end{revision}

Looking at the aggregate statistics, the FLN performs worst of all models analyzed; it is vastly outperformed by the RLN-t even though they have the same number of parameters.  \begin{revision} This is also reflected in the learning curve in Figure \ref{fig:learning_curve}: the RLN-t trained faster than the FLN and converged to a more accurate model simply by applying the proximal operator repeatedly across multiple iterations. Intriguingly, the FLN results are somewhat worse than the DL results. This implies that our simple network topology and relative lack of hyperparameter tuning produced a less-powerful model than the DL.
\end{revision}
However, the RLN-t did much better, producing less error on average (and significantly less variance) than the DL control for both contrast ratios. \begin{revision}
The disparity in performance indicates that dataset and network topology alone cannot explain the improved MASE relative to the DL model -- the iterative methodology produces a more powerful model.
\end{revision}

The full RLN is the most accurate model, producing roughly half as much error and variability as the DL model for both contrast ratios. \begin{revision}
The improvement over the RLN-t has a number of possible origins: the RLN uses a factor of $N$ more parameters, and in turn it uses a different operator at each iteration. We note that after training, the full RLN increases the memory overhead but not the prediction time: up to GPU memory details, each iteration has the same computational costs regardless of weight tying. Excluding data I/O overhead, the RLN-t and the full RLN take only\end{revision} $\approx 87$ seconds to predict strain fields for the entire testing set (8,192 microstructures), or roughly 11 milliseconds per microstructure (\textit{c.f.} $\approx$ 5 seconds per microstructure for FEA).

\begin{figure}
    \centering
    \includegraphics[width=\columnwidth]{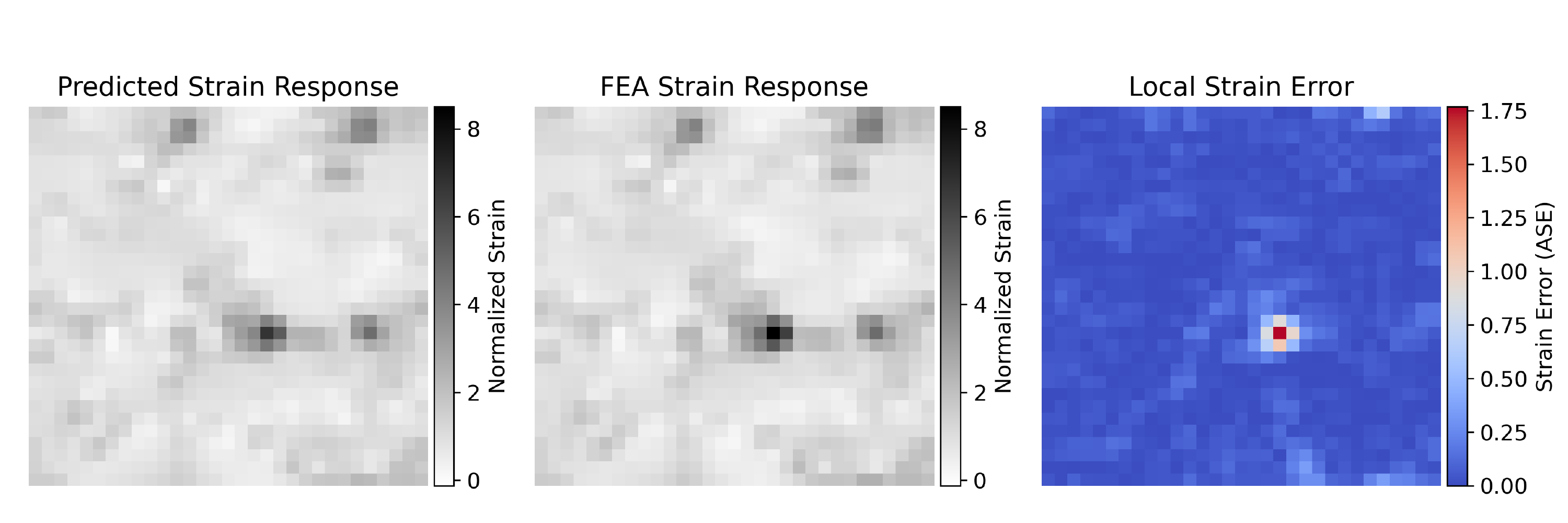}
    \caption{Slices of RLN predicted strain, true (FEA) strain, and ASE for worst test-set instance, $CR=50$}
    \label{fig:strain_rln}
\end{figure}

Figure \ref{fig:strain_rln} presents the worst (by voxel) test-set slice for the RLN model compared to the FEA-generated fields. The RLN appears to perform poorest near extreme strain peaks, especially in a 3-voxel wide cube around these peaks. This is caused by two issues. First, the microstructure is effectively undersampled in these areas. Referring back to Figure \ref{fig:micro_sample}, one sees that many microstructures have spatial features that are only one or two voxels wide. Furthermore, the output of the RLN is effectively `smeared' near these peaks due to the usage of a 3x3x3 filter. A deeper network, or one using multiple filter sizes, might be able to better capture these features. 

\begin{revision}

Finally, we explore how the predicted strain field $\bmy_i$ evolves across iterations, as well as its difference $\Delta_i \equiv \bmy_i - \bmy_{i-1}$. This is presented in Figure \ref{fig:evolution_fln}  for the FLN, Figure \ref{fig:evolution_rln_t} for the RLN-t, and Figure \ref{fig:evolution_rln} for the full RLN. The greyscale coloring is the same scale as Figure \ref{fig:strain_rln} and has its maximum at the true strain response's maximum. The redscale coloring represents a negative strain update ($\bmy$ decreasing between iterations). 

\begin{figure}
    \centering
    \includegraphics[width=0.35\columnwidth]{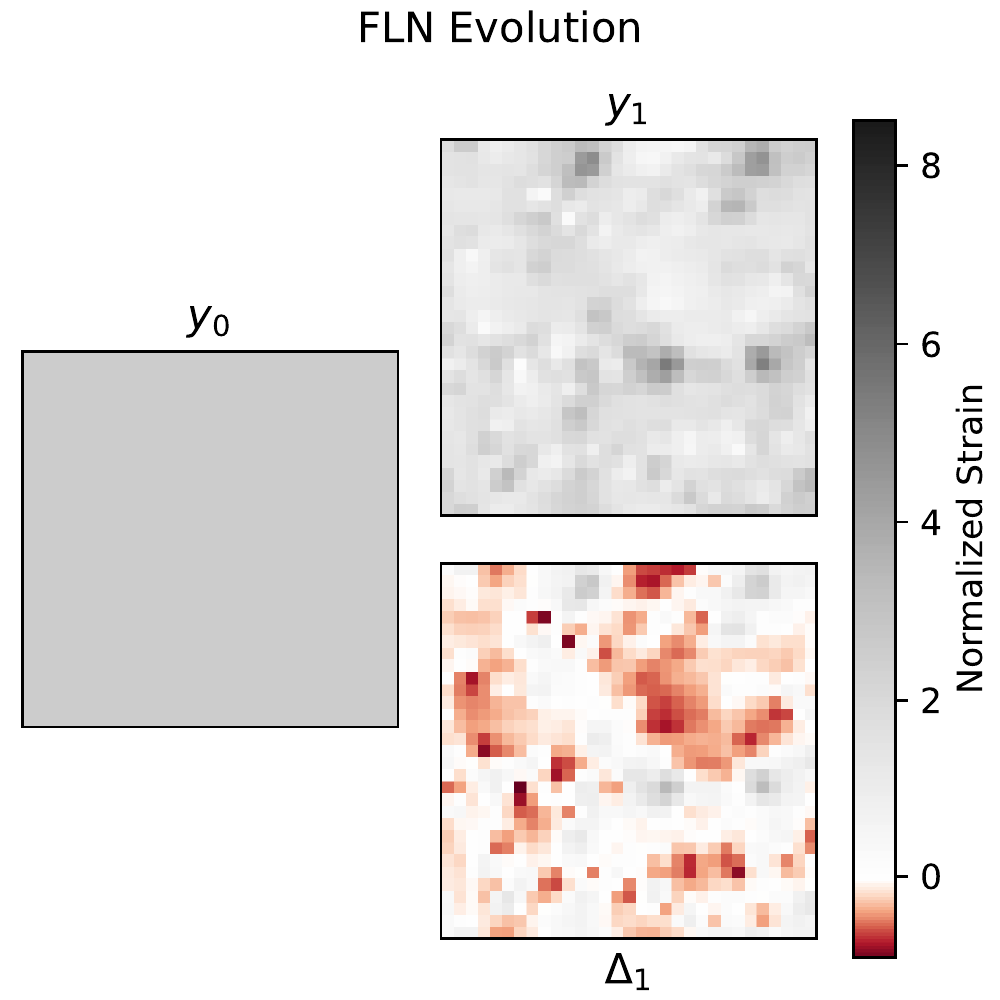}
    \caption{FLN predictions and differences between iterations for selected test-set instance, $CR=50$}
    \label{fig:evolution_fln}
\end{figure} 

\begin{figure}
    \centering
    \includegraphics[width=\columnwidth]{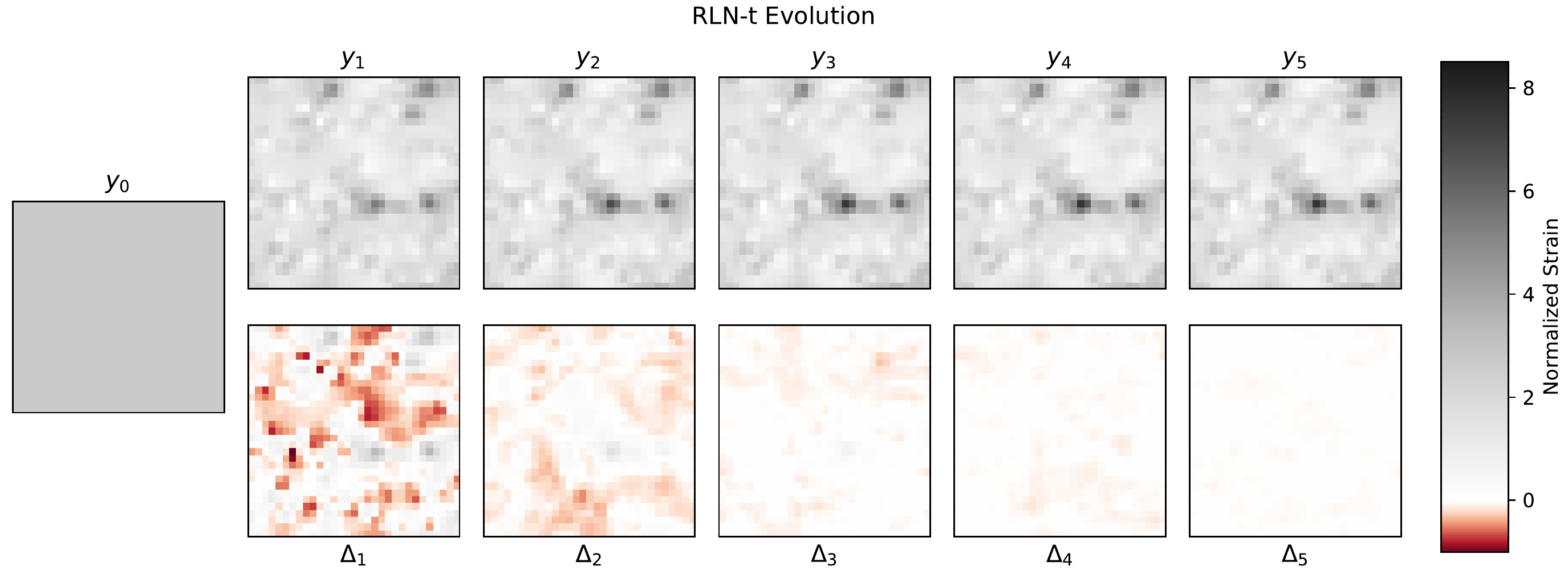}
    \caption{RLN-t predictions and differences between iterations for selected test-set instance, $CR=50$}
    \label{fig:evolution_rln_t}
\end{figure} 

\begin{figure}
    \centering
    \includegraphics[width=\columnwidth]{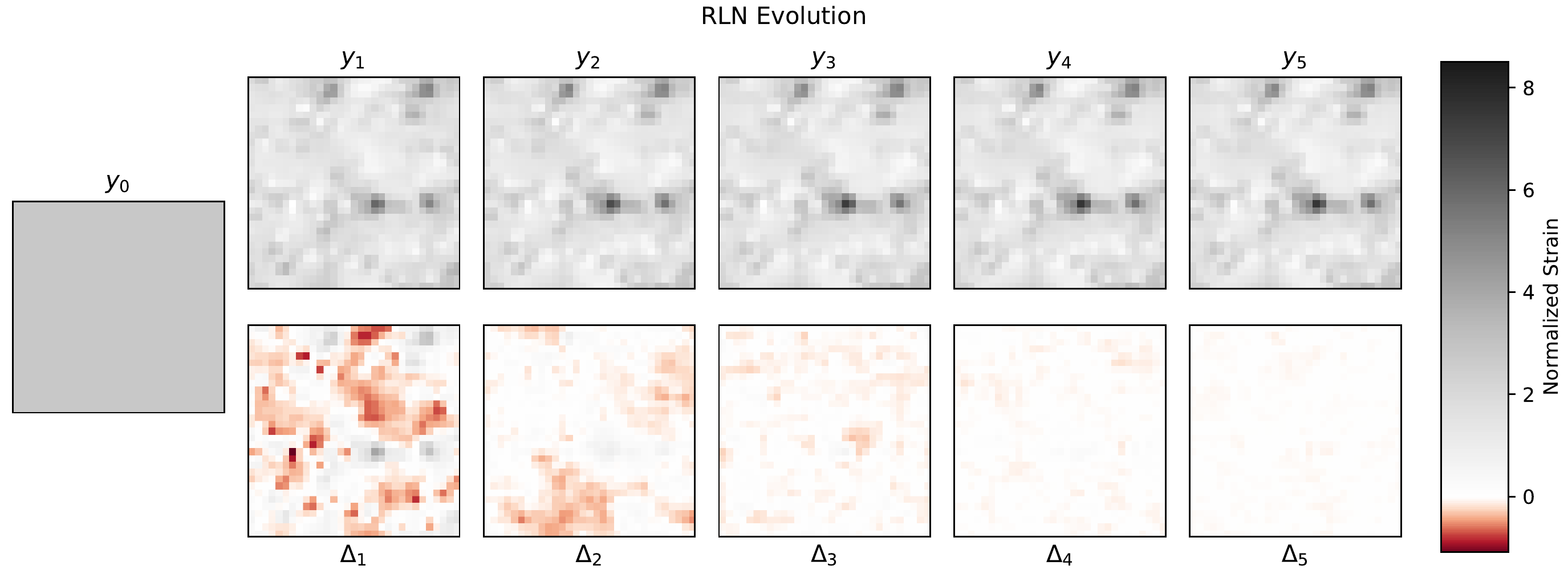}
    \caption{RLN predictions and differences between iterations for selected test-set instance, $CR=50$}
    \label{fig:evolution_rln}
\end{figure}

The FLN appears to handle low-strain areas fairly well, but misses the strain peaks in the center. Note that it does not apply the same operation as the first iteration of the RLN-t; the FLN is trained to make the best possible answer within a single jump. In comparison, the RLN-t does a better job of capturing strain peaks, likely since it can build them up over several iterations. What is less clear is that it fails to capture the central strain peak as well as the full RLN -- the differences between the two are evidently rather fine-scale, so we focus our analysis on the full RLN. 
\end{revision}

After the first iteration the RLN has picked up most of the relative troughs and peaks, and finer tweaks are handled by the later iterations. For example, most of the high-strain regions (greyscale areas) appear to be captured in the first two iterations, whereas low-strain regions (redscale areas) continue to be refined in later iterations.

This is due to the RLN's ability to learn different,  \begin{revision}and nonlinear,\end{revision} operators at each iteration -- the MSE loss function tends to magnify high-magnitude errors, even if they are very localized (i.e., strain peaks and troughs). The model is therefore encouraged to estimate these outlier areas first (corresponding to $\Delta_1$ having much more magnitude). Of course, this puts a lot of weight on the first proximal network to capture the strain outliers correctly. One possible solution would be to adjust the loss function weighting so that early iterations are encouraged to converge gradually towards a good solution, rather than quickly towards a decent one. In other words, by allowing earlier iterations to mispredict strain peaks somewhat, the model may be more likely to escape local minima and actually obtain higher final-iteration accuracy.

\begin {revision} 
This approach differs from traditional finite-element based approaches in that it seeks a solution by learning Green's functions (and derivatives thereof) of the governing equation, rather than solving the governing equation numerically. This provides an advantage in prediction speed by requiring a rather costly (but one-time) training overhead. The computational complexity of convolving a 3D field containing $S$ voxels with a 3D stencil containing $k$ voxels is $O(S k)$ (since each voxel in the first field must be multiplied by each voxel in the stencil). For a fixed network topology, $k$ is a constant; therefore the prediction runtime will increase linearly with the number of microstructure voxels. A spectral method using the Fast Fourier Transform will cost at least $O(S \log S)$, and any numerical methods (finite element or otherwise) employing linear solvers will likely be even more expensive. We reiterate that using GPU parallelization, the RLN requires on average 11 milliseconds to predict the strain field for a given microstructure, compared to several seconds for the finite element solver. This makes it very valuable for inverse problems, where the localization problem must be solved thousands or even millions of times in order to solve a higher-level metaproblem \cite{chen2020}.  Once trained for a specific governing equation (e.g. linear elasticity) and set of material properties (e.g., contrast ratio), the RLN methodology can be applied to any voxelized microstructure. Although we only tested a $31^3$ structure, in principle a model could be trained on one structure size and used on another (possibly with reduced accuracy); this is the subject of ongoing work. Note that the model must be trained anew to predict strains for a different contrast ratio. 
\end{revision}

\FloatBarrier

\section{Conclusions}\label{sec:conclusions}
In this paper, we describe a new learning-based methodology for addressing Lippmann-Schwinger type physics problems. Embedding recurrent CNNs into a learned optimization procedure provides for a flexible, but interpretable, ML model. The design of proximal networks is informed by problem-specific domain knowledge; that knowledge also provides a physical interpretation of what the model is learning. Furthermore, the partitioned and convolutional structure of this approach acts as a regularizer by enforcing underlying physical properties such as mean field values and spatial invariance. The iterative scheme allows for emulation of a deeper network, vastly increasing model robustness without increasing the parameterization space. If space allows, using a different network for each iteration further improves the model's expressiveness and accuracy. 

When applied to the elasticity localization problem \begin{revision} and using our dataset, our model produced much more accurate and interpretable results than previous deep learning models produced on similar datasets\end{revision}, while being much faster than analytical approaches. The CNN architecture used here was designed for simplicity, but could be improved with more advanced techniques such as inception modules \cite{szegedy_2014inception}, perceptual losses \cite{johnson2016_perceptuallosses}, Fourier layers \cite{li2020fourier}, or variational layers \cite{shridhar2019}. Moreover, many hyperparameters, such as number of iterations and loss function weighting, can be tuned on a problem-specific basis.

\section{Acknowledgements}\label{sec:acks}

This work was supported by NSF Graduate Research Fellowship DGE-1650044, and SK acknowledges support from NSF 2027105. We used the Hive cluster supported by NSF 1828187 and managed by PACE at Georgia Institute of Technology, USA. The authors thank Anne Hanna, Andreas Robertson, Nic Olsen, and Mady Veith for insightful discussions that helped shape this work. 

\section{Data and Software Availability}
The raw and processed data required to reproduce these findings are available to download from [dataset] \url{https://www.dropbox.com/sh/pma0npf1wr86n9i/AADFc7xWNOe6WilrJQbSHC8Va} \cite{dropbox_data}. \begin{revision} The implementation and training code for each RLN configuration, as well as trained models,\end{revision} are available under an open source license on Github (see Ref. \cite{github_code}).

\clearpage 
\bibliographystyle{elsarticle-num}
\bibliography{refs.bib}

\end{document}